\documentclass[twocolumn,aps,footinbib,superscriptaddress]{revtex4-2}
\usepackage{amsmath,amssymb,bm,graphicx,hyperref}
\usepackage{latexsym}
\usepackage{braket}
\usepackage{mathrsfs}
\usepackage{cancel}
\usepackage{color}
\usepackage{multirow}
\usepackage{mathtools}
\usepackage{mathrsfs}
\usepackage{comment}
\usepackage{cancel}
\renewcommand{\i}{{\rm i}}
\newcommand{\e}{{\rm e}}

\renewcommand{\det}{{\rm det}\;}
\newcommand{\sgn}{{\rm sgn}}
\renewcommand{\Re}{{\rm Re}\;}
\renewcommand{\Im}{{\rm Im}\;}

\usepackage{xcolor}
\hypersetup{
    colorlinks=true,
    citecolor=[rgb]{0,0,0.55},
    linkcolor=[rgb]{0,0,0.55},
    urlcolor=[rgb]{0,0,0.55},
}
\usepackage{xr}
\begin{document}
\title{Characterization of topological insulators based on the
electronic polarization with spiral boundary conditions}
 \author{Masaaki Nakamura}
\affiliation{
Department of Physics, Ehime University Bunkyo-cho 2-5, Matsuyama, Ehime
790-8577, Japan}
 \author{Shohei Masuda}
\affiliation{
Department of Physics, Ehime University Bunkyo-cho 2-5, Matsuyama, Ehime
790-8577, Japan}
 \author{Satoshi Nishimoto}
\affiliation{
Department of Physics, Technical University Dresden, 01069 Dresden,
Germany}
\affiliation{
Institute for Theoretical Solid State Physics, IFW Dresden, 01171
Dresden, Germany}
\date{\today}
\begin{abstract}
 We introduce the electronic polarization originally defined in
 one-dimensional lattice systems to characterize two-dimensional
 topological insulators. The main idea is to use spiral boundary
 conditions which sweep all lattice sites in one-dimensional order.  We
 find that the sign of the polarization changes at topological
 transition points of the two-dimensional Wilson-Dirac model (the
 lattice version of the Bernevig-Hughes-Zhang model) in the same way as
 in one-dimensional systems. Thus the polarization plays the role of
 ``order parameter'' to characterize the topological insulating state
 and enables us to study topological phases in different dimensions in a
 unified way.
\end{abstract}

\maketitle

{\it Introduction.}
For more than ten years, topological phases and topological transitions
have been extensively studied in connection with topological insulators
\cite{Haldane1988,Kane-M,Bernevig-Z,Bernevig-H-Z2006,Ryu-S-F-L,
Konig-2007,Konig-2008}.  Topological insulators have energy gaps in the
bulk and gapless edge (surface) states in two (three) dimensions.
On the other hand, topological phases and topological transitions have
also been discussed in two-dimensional (2D) classical spin systems and
one-dimensional (1D) quantum spin systems since the 1970s
\cite{Berezinskii1971,Berezinskii1972,Kosterlitz-T}.  For example a
dimer-N\'eel transition in a spin-$1/2$ frustrated anisotropic
Heisenberg chain is regarded as a transition between two topologically
distinct gapped phases \cite{Haldane1982}. The discovery of the Haldane
gap in integer spin chains has added to the variety of topological
phases \cite{Haldane1983a,Haldane1983b}.
In this Research Letter, we study these topological phases and
topological transitions of different systems in different dimensions in
a unified way.

For this purpose, we consider the electronic polarization
\cite{Resta1994,Resta,Resta-S1999,Resta2000,Aligia-O}.  In 1D lattice
electron systems, the polarization operator is defined as the following
ground-state $\ket{\Psi_0}$ expectation value of the ``twist operator''
$U$,
\begin{equation}
 z^{(q)}=\braket{\Psi_0|U^q|\Psi_0},\quad
  U=\exp\biggl(\i\frac{2\pi}{L}\sum_{j=1}^L j n_j\biggr)
  \label{def_z}
\end{equation}
where $L$ is the number of sites, $n_j$ is the electron number operator
at $j$ th site, and $q$ is the degeneracy of the ground state.  Resta
related $z^{(1)}$ with the electronic polarization as
$\lim_{L\to\infty}(e/2\pi)\Im\!\ln z^{(1)}$ \cite{Resta}.  This quantity
$z^{(q)}$ has been calculated for several 1D systems
\cite{Resta-S1999,Nakamura-V,Nakamura-T}. Hereinafter we call $z^{(q)}$
itself ``polarization.'' The signs of $z^{(q)}$ identify topologies of
the systems such as charge or spin density waves.  By replacing $n_j$
with a spin operator $S_j^z$, $z^{(q)}$ can also identify several
magnetic orders including valence bond solid states
\cite{Nakamura-T}. Furthermore the condition $z^{(q)}=0$ can be used to
detect a phase transition point.

The same quantity as in Eq.~(\ref{def_z}) was also introduced in the
Lieb-Schultz-Mattis (LSM) theorem for 1D quantum systems
\cite{Lieb-S-M,Affleck-L,Affleck,Oshikawa-Y-A,Yamanaka-O-A}.  In the LSM
theorem, Eq.~(\ref{def_z}) appears as an overlap between the ground
state and a variational excited state. According to the LSM theorem, an
energy gap above a $q$-fold degenerate ground state is possible for
$z^{(q)}\neq 0$ with $L\to\infty$.

\begin{figure*}[t]
 \includegraphics[width=180mm]{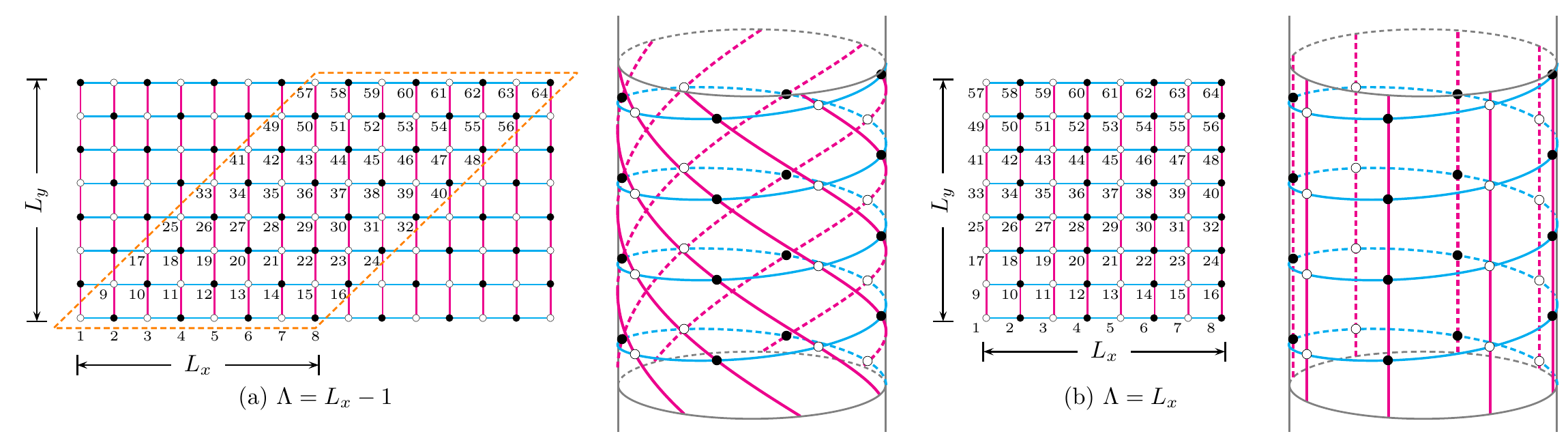}
 \caption{(a) and (b) Spiral boundary conditions (SBCs) for 2D square
 lattices where the systems are represented as extended 1D chains (blue
 lines). The parameter $\Lambda$ is the hopping range
 $c_{i+\Lambda,\alpha}^{\dag}c_{i,\alpha}^{\mathstrut}+\mbox{H.c.}$ of
 the 1D chain originating from the hopping along the $y$ direction
 (magenta lines). For even $L_x$, $\bm{k}=(\pi,\pi)$ [$\bm{k}=(\pi,0)$]
 order is represented by 1D $k=\pi$ modulation with $\Lambda=L_x-1$
 ($\Lambda=L_x$).  For odd $L_z$, the roles of $\Lambda$ are
 interchanged.}\label{fig:SBC}
\end{figure*}

Thus the property of the polarization $z^{(q)}$ has been well studied
for 1D systems, however, its application to higher dimensional systems
is not fully understood. In this Research Letter, we extend the twist
operator in Eq.~(\ref{def_z}) to 2D systems, characterize the
topological orders, and identify topological transition points.  The
main idea of our study is to use spiral boundary conditions (SBC) which
sweep all lattice sites in one-dimensional orders. For a 2D square
lattice with the number of lattices $L_x\times L_y$, SBC are introduced
as shown in Fig.~\ref{fig:SBC}.  These boundary conditions have been
introduced in extending the LSM theorem to higher dimensions
\cite{Oshikawa,Hastings} to remove unphysical restrictions for the
system sizes \cite{Yao-O}. We further introduce the parameter $\Lambda$
to deal with a variety of modulations. Then we show that topological
insulating states in 2D systems can be identified by the polarization
(\ref{def_z}) with SBCs. Throughout this Research Letter, the lattice
constant $a$ and the reduced Planck constant $\hbar$ are set to unity.

{\it The Wilson-Dirac model.}
As a fundamental model to describe 2D topological insulators, we
consider the Wilson-Dirac model \cite{Wilson,Qi-W-Z}, which is the
lattice version of the Bernevig-Hughes-Zhang (BHZ) model
\cite{Bernevig-Z,Bernevig-H-Z2006},
\begin{subequations} 
 \label{WD}
\begin{align}
 \mathcal{H}
  =&\sum_{\bm{k},\alpha, \beta} c^\dag_{\bm{k}, \alpha}
  \,H_{\alpha \beta}(\bm{k}) \,c^{\mathstrut}_{\bm{k}, \beta},
 \label{WD.1}\\
  H(\bm{k})
  =&t\sum_{\mu =x, y}\sin k_\mu \,\tau_\mu
 +\left[M - B \sum_{\mu = x, y}
 \bigl(1 -\cos k_\mu \bigr)\right]\tau_z,
 \label{WD.2}
\end{align}
\end{subequations}
where $t$ is the hopping amplitude, $M$ is the mass, $B$ is the
coefficient of the Wilson term, $c_{\bm{k},\alpha}$ is the annihilation
operator of a fermion with a 2D wave number, $\alpha,\beta$ are orbital
indices, and $\tau_{\mu}$ are the Pauli matrices. The energy eigenvalue
is given by
\begin{equation}
 \varepsilon_{\bm{k}}^2=t^2(\sin^2 k_x+\sin^2 k_y)
  +\bigl\{M - B(2-\cos k_x -\cos k_y)\bigr\}^2.
  \label{sxy_continuum}
\end{equation}
This system is a topological (trivial) insulator for $B<M/4$ ($B>M/4$),
and a phase transition among two topological phases occurs at $B=M/2$.
These transition points can be identified by vanishing of the bulk
energy gap $\varepsilon_{\bm{k}}=0$.  In the continuum version of the
model, the Hall conductivity is calculated as \cite{So}
\begin{equation}
  \sigma_{xy}=-\frac{e^2}{2h}[\sgn(M)+\sgn(B)].
\end{equation}
Therefore the system is a topological (trivial) insulator for $MB>0$
($MB<0$), and a topological transition occurs at $B=0$ for fixed $M$.

Now we consider the lattice model (\ref{WD}) based on SBC.  Here SBC
are introduced by replacing the 2D wave vector $\bm{k}=(k_x,k_y)$ as $k_x\to
k$ and $k_y\to\Lambda k$ with the 1D Fourier transformation
\begin{equation}
 c_{k,\alpha}=\frac{1}{\sqrt{L}}\sum_{j=1}^L\e^{-\i k x_j}c_{j,\alpha},
\end{equation}
where $L\equiv L_xL_y$, $x_j=aj$, and $k=2\pi n/L$ with
$n=0,1,2,\cdots,L-1$ meaning periodic boundary conditions (PBCs) for the
extended 1D chain, $c_{i+L,\alpha}=c_{i,\alpha}$.  When $L_x$ is even,
the parameter $\Lambda$ is chosen as $\Lambda=L_x-1$ [$\Lambda=L_x$] to
detect a modulation of $\bm{k}=(\pi,\pi)$ [$\bm{k}=(0,\pi),(\pi,0)$] as
shown in Fig.~\ref{fig:SBC}. The present system has translational
symmetry $\mathcal{T}c_{j,\alpha}\mathcal{T}^{-1}=c_{j+1,\alpha}$ and
parity symmetry $\mathcal{P}c_{j,\alpha}\mathcal{P}=c_{L-j+1,\alpha}$,
so that $z^{(q)}=z^{(q)}\e^{-\i2q\pi N/L}$ and
$z^{(q)}=[z^{(q)}]^*\e^{\i2q\pi N/L}$ with $N$ being the number of
fermions.  Thus we should choose $q=1$ in the present case with $N=L$.

{\it Polarization.}
In order to calculate the polarization $z^{(1)}$ for the 2D Wilson-Dirac
model, we use the following Resta's argument \cite{Resta}.  After the
inverse Fourier transformation, the Wilson-Dirac model (\ref{WD}) with
SBCs is written as a 1D quadratic Hamiltonian \cite{SM},
\begin{equation}
 \mathcal{H} = \sum_{ij,\alpha\beta}
 c_{i\alpha}^\dagger H_{ij,\alpha \beta}c_{j\beta}^{\mathstrut},
\end{equation}
where $i,j$ are sites.  Then we obtain its single-particle eigenstates
by
\begin{equation}
 \sum_{j'}H_{jj'}\ket{\psi^{j'}_{p\mu}}
  =\varepsilon_{p\mu}\ket{\psi^{j}_{p\mu}},
\end{equation}
where $\ket{\psi^{j}_{p\mu}}=\mathcal{U}^{-1}_{jp}\ket{u_{p\mu}}$,
$\mathcal{U}^{\mathstrut}_{pj}=\e^{-\i p x_j}/\sqrt{L}$, and
$\ket{u_{p\mu}}$ are the eigenstates of the Bloch Hamiltonian
$H(p)=\sum_{jj'}\mathcal{U}^{\mathstrut}_{pj}H_{jj'}\mathcal{U}^{-1}_{j'p}$.
Then the polarization is given as
\begin{equation}
 z^{(q)}={\rm det}' S^{(q)},\quad S^{(q)}_{\mu,\nu}(k_{s},k_{s'})
  =\sum_{j=1}^{L}\braket{\psi^{j}_{k_{s}\mu}|
  \e^{\i\frac{2q\pi}{L}x_j}|\psi^{j}_{k_{s'}\nu}},
\end{equation}
where det$'$ indicates the determinant restricted to the occupied
single-particle states. This calculation is simplified as
\begin{align}
 &\det S^{(q)}=(-1)^{q}\prod_{s=0}^{L-1}\det S^{(q)}(k_{s+q},k_{s}),\\
 &S^{(q)}_{\mu,\nu}(k_{s+q},k_s)=
 \braket{u_{k_{s+q}\mu}|u_{k_{s}\nu}},
\end{align}
where $k_s=2\pi s/L$, and the factor $(-1)^{q}$ stems from the
antisymmetry of the determinant \cite{SM}.  Here the meaning of the
polarization becomes clear: it is a product of overlaps between Bloch
states with wave vectors that differ by $k_q$.  This decomposition to
small matrices and cancellation of $x_j$ dependence greatly simplify the
calculations and enable us to deal with large systems. Especially, in
the present system with a single occupied band,
$S_{\mu,\nu}(k_{s+1},k_s)$ is no longer a matrix but a number.  This
simplification is also one of the advantages of SBCs where each state is
specified by a single wave number $k$, compared with conventional 2D
PBCs.

{\it Conductivities.}
As a physical quantity to compare with the polarization, we calculate
the Hall conductivity $\sigma_{xy}$ which is given in Matsubara form as
\begin{subequations} 
\begin{align}
 &\Re\sigma_{ij}
 =-\lim_{\omega\to 0}
 \frac{\Im\Pi_{ij}^{\rm R}(0,\omega)}{\omega}
 \qquad \{i,j\}\in\{x,y\}\\
 &\Pi_{ij}(p,\i\nu_m)=
 \frac{e^2}{L_{i}L_{j}\beta}
 \sum_{k,\omega_n}\\
 &\times
 {\rm Tr}\left[\mathcal{G}(k,\i\omega_n)
 \,\gamma_{i}\left(k+\tfrac{p}{2}\right)
 \mathcal{G}(k+p,\i\omega_n+\i\nu_m)
 \,\gamma_{j}\left(k+\tfrac{p}{2}\right)\right] \nonumber
\end{align}
\end{subequations}
where $\beta$ is the inverse temperature, and the temperature Green's
function is given as $\mathcal{G}(k,\i\omega_n)
=(\i\omega_n-(H(k)-\mu)+\i\,\sgn(\omega_n)\Gamma)^{-1}$ with
$\omega_{n}$ and $\nu_m$ being Matsubara frequencies for fermions and
bosons, respectively.  The chemical potential and the impurity
scattering time are denoted by $\mu(=0)$ and $1/2\Gamma$,
respectively. $\gamma_i$ is defined by
\begin{equation}
 \gamma_{i}(\bm{k})=\frac{\partial H(\bm{k})}{\partial k_i},
\end{equation}
and the replacement of the wave number $(k_x,k_y)\to(k,\Lambda k)$
\cite{SM}.

\begin{figure}[t]
 \includegraphics[width=80mm]{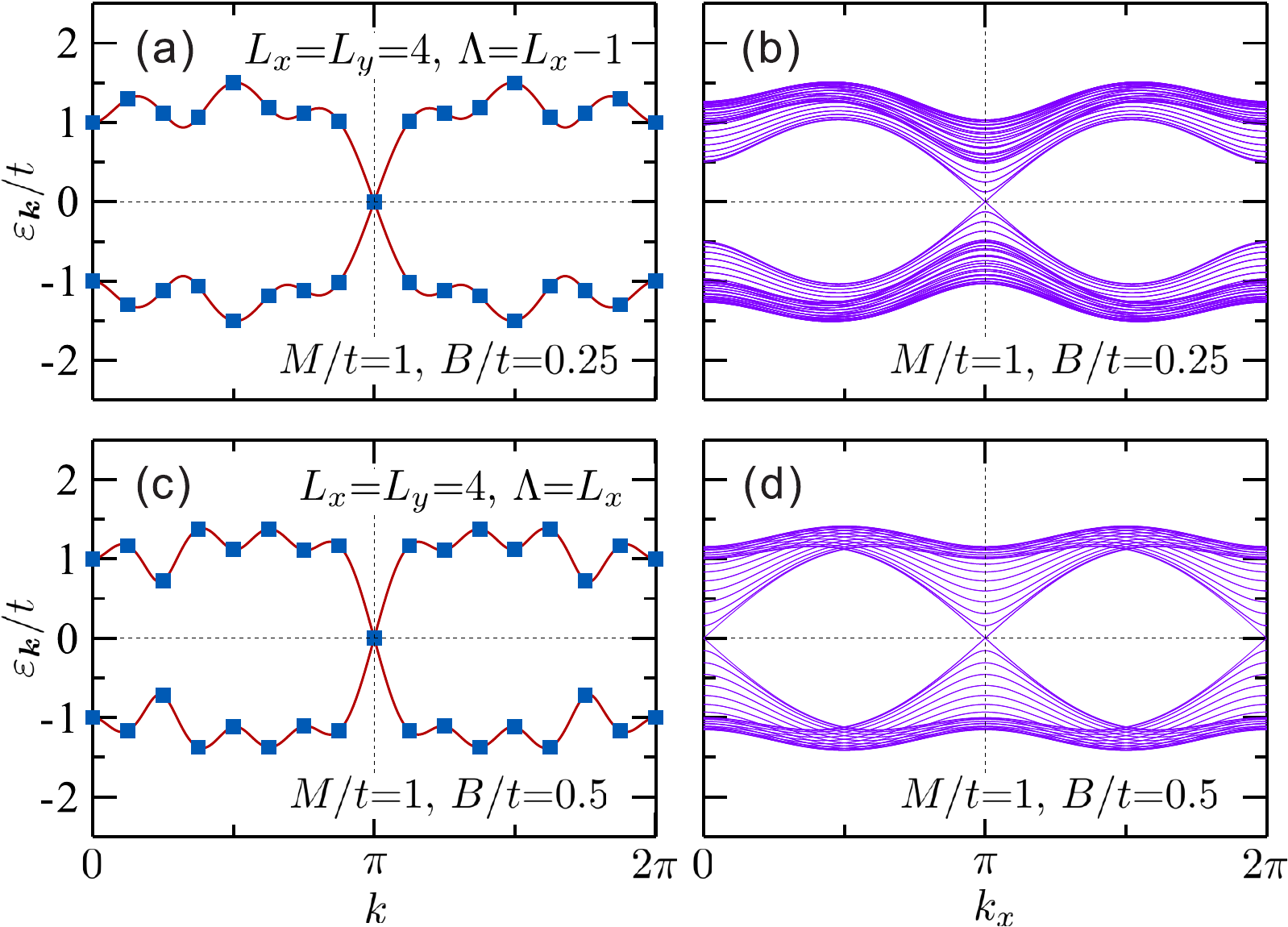}
 \caption{Dispersion relations of the 2D Wilson-Dirac model
 ($L_x=L_y=4$, $M/t=1$) at $B=M/4$ [(a) and (b)] and at $B=M/2$ [(c) and
 (d)].  (a) and (c) are written with spiral boundary conditions for
 $\Lambda=L_x-1$ and for $\Lambda=L_x$, respectively, while (b) and (d)
 are based on conventional 2D boundary conditions.  The bulk gap closes
 and the Dirac dispersion appears at $k=\pi$.}  \label{fig:dispersion}
\end{figure}

{\it Results.}
First, we look at the dispersion relation of the 2D Wilson-Dirac model
with SBCs as shown in Fig.~\ref{fig:dispersion}.  The Brillouin zone is
$0\leq k <2\pi$. Since the present model with SBCs is represented as a
1D model with long-range hopping terms
$c_{i+\Lambda,\alpha}^{\dag}c_{i,\alpha}^{\mathstrut}+\mbox{H.c.}$,
there appear many oscillations. At the phase transition points where the
bulk gap closes, Dirac points appear at $k=\pi$ for $\Lambda=L_x-1$ and
$\Lambda=L_x$. Here we have assumed that $L_x$ is even. For odd $L_x$,
the roles of $\Lambda$ are interchanged.

\begin{figure}[t]
 \includegraphics[width=80mm]{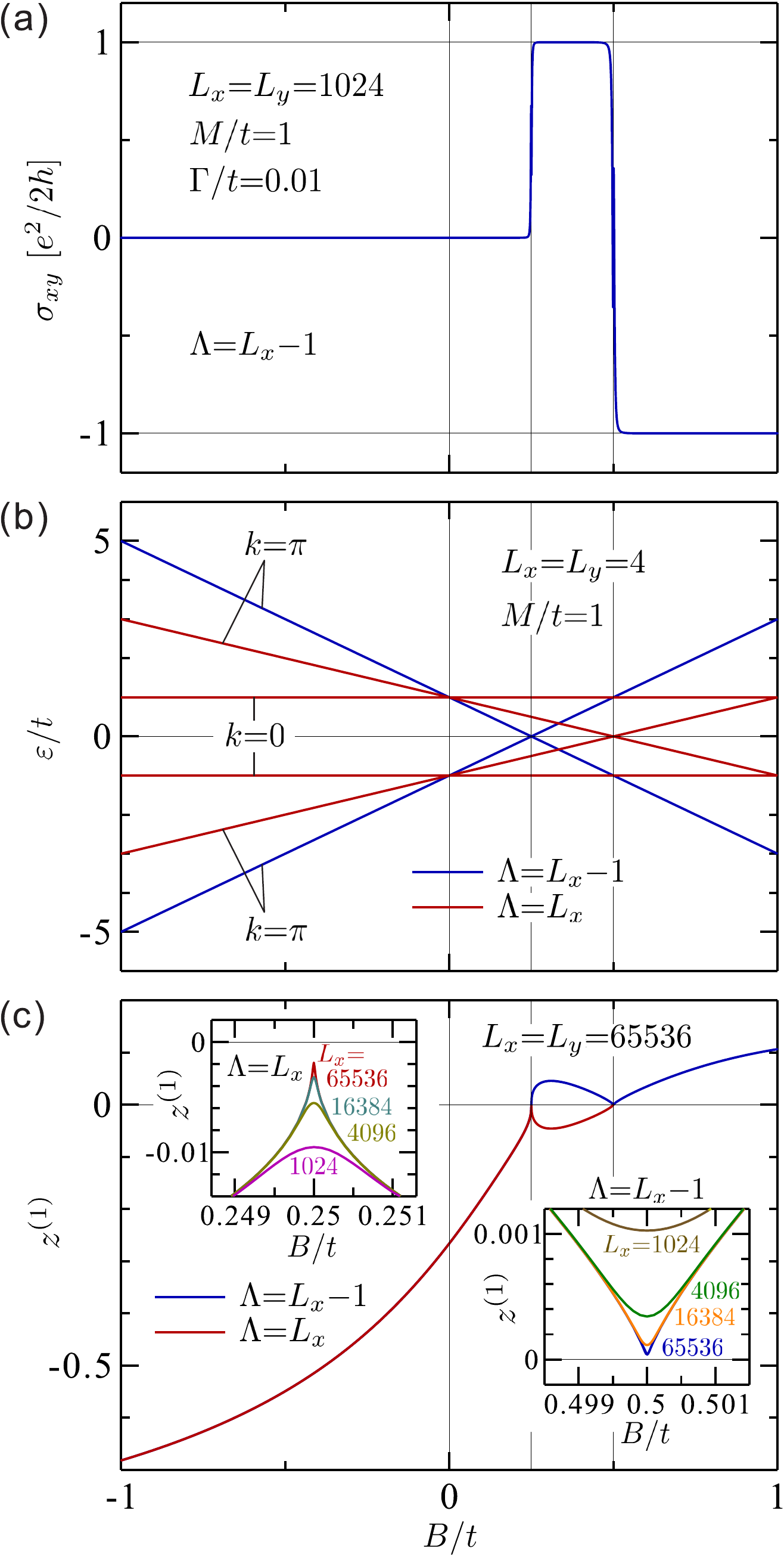}
 \caption{Several quantities around the parameter regions of the
 topological transitions of the 2D Wilson-Dirac model with $M/t=1$: (a)
 Hall conductivity, (b) energy levels in a finite-size system, and (c)
 polarization. The Hall conductivity is quantized at $\pm e^2/2h$. (a)
 and (c) are calculated under SBC with APBC
 $c_{i+L,\alpha}=-c_{i,\alpha}$ to prevent divergences at the transition
 points ($B/t=1/4, 1/2$). The insets of (c) show how $z^{(1)}$
 approaches to zero at the transition points. The polarizations with
 different $\Lambda$ converge to the same value except for the
 intermediate region $1/4<B/t<1/2$.}\label{fig:data}
\end{figure}

We calculate several quantities including the polarization as shown in
Fig.~\ref{fig:data}. As expected, the Hall conductivity at zero
temperature vanishes ($\sigma_{xy}=0$) in the trivial phase $B<M/4$ and
$\sigma_{xy}=\pm e^2/2h$ for the topological phase $B>M/4$ as shown in
Fig.~\ref{fig:data}(a). 
This is due to the absence of chiral symmetry.
At the phase transition between two topological states $B=M/2$, the sign
of the Hall conductivity changes.  The results of the lattice model do
not coincide with those of the effective mass approximation.  Since the
Hall conductivity is defined in the thermodynamic limit, the results do
not depend on boundary conditions: we get the same quantized Hall
conductivities for SBC with $\Lambda=L_x-1$ and $\Lambda=L_x$, and also
for the usual 2D PBCs.  However, at the phase transition points, the
Hall conductivity diverges because of the vanishing of the bulk energy
gap $\varepsilon_{k}=0$. Therefore we have calculated the Hall
conductivity in antiperiodic boundary conditions (APBCs) for the
extended 1D chain, $c_{i+L,\alpha}=-c_{i,\alpha}$, namely
$k=2\pi(n+1/2)/L$, to prevent these divergences.

Next we consider energy spectra of the 2D Wilson-Dirac model in
finite-size systems. As shown in Fig.~\ref{fig:data}(b), energies of
$\bm{k}=(\pi,\pi)$ $[(k,\Lambda)=(\pi,L_x-1)]$ intersect at the
topological phase transition point at $B=M/4$ without size dependence.
This is similar to ``level spectroscopy'' for 1D quantum systems
\cite{Okamoto-N,Nomura-O,Kitazawa}. In this method, phase transition
points between two different gapped states described by $c=1$ conformal
field theory are identified by an intersection of energy spectra with
different parities. The phase transition point at $B=M/2$ is also given
by an intersection of energy spectra, but these spectra are given by
$\bm{k}=(\pi,0)$ $[(k,\Lambda)=(\pi,L_x)]$, so that we need to choose
$\Lambda=L_x$.

Now we turn our attention to the polarization $z^{(1)}$ which is the
main target of this Research Letter.  Before looking at the result, let
us evaluate the sign of $z^{(1)}$ from the real space representation of
the 2D Wilson-Dirac model \cite{SM}. For the trivial phase ($B\ll 0\ll
M$), the electrons of the system are located on each site; therefore it
follows from $(2\pi/L)\sum_{j=1}^L j=(L+1)\pi$ that $z^{(1)}=-1$ for
even $L$. On the other hand, the topological phase ($0\ll M\ll B$) is
considered to be dominated by bond-located fermions. When all the
fermions are located on the bonds along the $x$ direction, the wave
function is $\ket{\Psi_0}=2^{-L/2}\prod_{k=1}^{L/2}
(c_{2k-1,1}^{\dag}+c_{2k,1}^{\dag})
(c_{2k,2}^{\dag}+c_{2k+1,2}^{\dag})\ket{0}$.
Then the polarization is calculated as
$z^{(1)}=[\cos(\pi/L)]^L\e^{\i(L+2)\pi}\to 1$ for $L\to\infty$. Here,
the phase is given by the center of mass of the bond-located fermions
$(2\pi/L)\sum_{j=1}^L (j+1/2)$. If there are bond-located fermions along
the $y$ direction, then $\ket{\Psi_0}$ includes long range bonds
$2^{-1/2}(c_{j,\alpha}^{\dag}+c_{j+\Lambda,\alpha}^{\dag})\ket{0}$ in
the 1D representation. However, in such cases, the center of mass of the
bond-located fermions is unchanged regardless of the choice of $\Lambda$
\cite{SM}.  Therefore, the sign of $z^{(1)}$ is expected to change
between the topological and the trivial phases.

As shown in Fig.~\ref{fig:data}(c), the sign of $z^{(1)}$ changes
between the trivial and the topological regions, as we expected.  For
the region between two phase transition points at $B=M/4$ and $B=M/2$,
the sign of $z^{(1)}$ becomes different depending on the choice of
$\Lambda$. The interpretation of this result is not clear, but the
difference in the sign of $z^{(1)}$ indicates the system given by
bond-located fermions with modulations characterized by
$(k_x,k_y)=(\pi,\pi)$ or $(\pi,0),(0,\pi)$.  For even $L_x$, $z^{(1)}$
with $\Lambda=L_x-1$ changes the sign at $B=M/4$ and approaches to zero
at $B=M/2$, while for $\Lambda=L_x$, $z^{(1)}$ approaches to zero at
$B=M/4$ and changes the sign at $B=M/2$. The fact that $z^{(1)}$
vanishes at the phase transition points where the system is gapless is
consistent with the LSM theorem. Here, we have used SBCs with APBCs
$c_{i+L,\alpha}=-c_{i,\alpha}$ to prevent divergences at the transition
points. For PBCs, $z^{(1)}$ changes discontinuously at the
level-crossing point in finite-size systems \cite{Nakamura-F}.

{\it Summary and discussion.}
In summary, we have discussed the polarization in the 2D Wilson-Dirac
model based on spiral boundary conditions that sweep all lattice sites
in 1D order.  Here the system is described as 1D chains with long-range
hopping. Then the electronic polarization defined in 1D systems can be
extended to 2D systems. In the same way as in the 1D cases,
topologically distinct gapped phases are characterized by the difference
of the sign in the polarization. This means that the polarization
operator and SBCs enable us to deal with topological transitions between
different gapped states in different dimensions in a unified way.

%
%
The SBCs also have a great advantage in calculating the polarization.
In Resta's formalism for non-interacting systems, the polarization
$z^{(q)}$ is given by products of overlaps between the Bloch states with
wave vectors separated by $2q\pi/L$, so that denoting the states by 1D
wave numbers greatly reduces the calculation costs, and enables us to
obtain results in large enough systems to be almost regarded as the
thermodynamic limit.  The present SBCs are also useful for several
numerical methods, such as the exact diagonalization and the density
matrix renormalization group.

%
%
We have analyzed the 2D Wilson-Dirac model successfully based on SBC,
but we can not conclude here whether the behavior of the polarization is
universal or not in other models of 2D topological insulators.  We need
further systematic study of the general relationship between the
polarization and topological phases in several symmetries and
dimensions, including calculations of other physical quantities such as
entanglement spectra.
For example, there are several works relating the Chern number and the
conventional type of 2D twist operators \cite{Coh-V,Kang-L-C}.
It would also be interesting to apply the present analysis to multipole
polarizations in higher-order topological insulators
\cite{Kang-S-C,Wheeler-W-H,Watanabe-O} and non-Hermitian systems.

{\it Acknowledgments.}
The authors thank K.~Imura and M.~Oshikawa for discussions, and
U.~Nitzsche for technical assistance.  M. N. acknowledges the Visiting
Researcher's Program of the Institute for Solid State Physics, The
University of Tokyo, and the research fellow position of the Institute
of Industrial Science, The University of Tokyo.  M.~N. is supported
partly by MEXT/JSPS KAKENHI Grant Number JP20K03769. S.~N. acknowledges
support from SFB 1143 project A05 (project-id 247310070) of the Deutsche
Forschungsgemeinschaft.


\end{document}